\documentclass[reprint,longbibliography,aps,prl]{revtex4-1}
\usepackage{mathtools,amssymb,amsthm}
\usepackage[utf8]{inputenc}
\usepackage{graphicx}
\usepackage{hyperref}

\newcommand{\be}{\begin{equation}}
\newcommand{\ee}{\end{equation}}
\newcommand{\bi}{\begin{itemize}}
\newcommand{\ei}{\end{itemize}}
\def\ba#1\ea{\begin{align}#1\end{align}}
\def\bg#1\eg{\begin{gather}#1\end{gather}}
\def\bm#1\em{\begin{multline}#1\end{multline}}
\def\bmd#1\emd{\begin{multlined}#1\end{multlined}}

\def\b{\beta}

\def\d{\delta}

\def\e{\epsilon}

\def\r{\rho}

\def\S{\Sigma}

\newcommand{\la}{\label}

\newcommand{\re}{\ref}
\newcommand{\er}{\eqref}
\newcommand{\se}{\section}

\newcommand{\fr}{\frac}

\newcommand{\pa}{\partial}

\newcommand{\wtd}{\widetilde}

\newcommand{\eq}{\equiv}

\renewcommand{\ol}{\overline}

\renewcommand{\(}{\left(}
\renewcommand{\)}{\right)}
\renewcommand{\[}{\left[}
\renewcommand{\]}{\right]}

\renewcommand{\>}{\rangle}

\newcommand{\tr}{\operatorname{tr}}

\newcommand{\bZ}{{\mathbb Z}}

\newcommand{\cB}{{\mathcal B}}
\newcommand{\cC}{{\mathcal C}}
\newcommand{\cE}{{\mathcal E}}

\newcommand{\Area}{\operatorname{Area}}

\newcommand{\mc}{\text{mc}}
\renewcommand{\th}{\text{th}}

\begin{document}

\title{Holographic R\'enyi Entropy at High Energy Density}
\author{Xi Dong}
\email{xidong@ucsb.edu}
\affiliation{Department of Physics, University of California, Santa Barbara, California 93106, USA}

\begin{abstract}
We show that R\'enyi entropies of subregions can be used to distinguish when the entire system is in a microcanonical ensemble from when it is in a canonical ensemble, at least in theories holographically dual to gravity.  Simple expressions are provided for these R\'enyi entropies in a particular thermodynamic limit with high energy density and fixed fractional size of the subregion.  Holographically, the R\'enyi entropies are determined by the areas of cosmic branes inserted into the bulk spacetime.  They differ between a microcanonical and a canonical ensemble because the two ensembles provide different boundary conditions for the gravitational theory under which cosmic branes lead to different backreacted geometries.  This is in contrast to the von Neumann entropy which is more coarse-grained and does not differentiate microcanonical ensembles from canonical ensembles.
\end{abstract}

\maketitle

\se{Introduction}

In a chaotic quantum system, the eigenstate thermalization hypothesis~\cite{Deutsch1991quantum,Srednicki1994chaos,Rigol2008thermalization} posits that a finitely excited energy eigenstate behaves thermally when restricted to a subsystem much smaller than the entire system.  In other words, few-body operators cannot distinguish an energy eigenstate from a suitable thermal state.

This raises an important question: are there more \textit{global} probes that can differentiate an energy eigenstate from a thermal state?  In particular, it is well motivated to consider probes defined in a subsystem $A$ whose fractional size $f \eq V_A/V$ is finite in the thermodynamic limit.  As a first step in studying this question, it is useful to address a related question: what probes on a subsystem can be used to distinguish when the entire system is in a microcanonical ensemble at fixed total energy from when it is in a canonical ensemble at an appropriate temperature?

The main purpose of this Letter is to answer this second question in holographic theories by showing that the R\'enyi entropies~\cite{Renyi:1961}
\be
S_n^A \eq \fr{1}{1-n} \ln \tr \r_A^n
\ee
of any subsystem $A$ of finite fractional size differ between a microcanonical and a canonical ensemble for the entire system as long as $n\neq1$.  This suggests that, even when restricted to a subsystem, R\'enyi entropies have much more fine-grained information about the entire state than the von Neumann entropy.  By a happy coincidence, R\'enyi entropies (for integer $n\ge2$) are much easier to measure experimentally than the von Neumann entropy, as recently demonstrated using ultracold atoms~\cite{Islam:2015measuring}.  Moreover, the results of this Letter are motivated by similar observations made recently in Ref.~\cite{Lu:2017tbo} (which is itself motivated by Refs.~\cite{Garrison:2015lva,Dymarsky:2016ntg}) for individual energy eigenstates instead of microcanonical ensembles, using arguments based on ergodicity and eigenstate thermalization rather than holography.

Our main technical tool is a simple, geometric prescription derived in Refs.~\cite{Lewkowycz:2013nqa,Dong:2016fnf} for calculating R\'enyi entropies using gauge-gravity duality~\cite{Maldacena:1997re,Gubser:1998bc,Witten:1998qj}.  It says that in quantum field theories (QFTs) with a dual gravitational description in a bulk spacetime with one additional dimension, a refined version of the R\'enyi entropy introduced in Ref.~\cite{Dong:2016fnf},
\be\la{tsn}
\wtd{S}_n^A \eq n^2 \pa_n \(\fr{n-1}{n} S_n^A\) = -n^2 \pa_n \(\fr{1}{n} \ln \tr \r_A^n\),
\ee
is given by the area of a bulk codimension-2 cosmic brane homologous to the subregion $A$:
\be\la{tsnb}
\wtd{S}_n^A = \fr{\Area(\text{Cosmic Brane}_n)}{4 G_N}.
\ee
The bulk geometry satisfies the equations of motion (Einstein's equations).  In particular, the cosmic brane backreacts on the bulk geometry by creating a conical defect with an opening angle $2\pi/n$ because of its tension $(n-1)/(4n G_N)$.  Once we know the refined R\'enyi entropies $\wtd{S}_n^A$, the R\'enyi entropy $S_n^A$ for any positive $n$ is easily obtained by integrating~\er{tsn}:
\be\la{sni}
S_n^A = \fr{n}{n-1} \int_1^n \fr{\wtd{S}_{n'}^A}{n'^2} dn'.
\ee

This holographic prescription for R\'enyi entropies works in the large $C$ limit of the boundary QFT where $C$ is a generalized ``central charge'' describing the number of degrees of freedom per short-distance cutoff. In this limit we are instructed to determine a bulk solution with a suitable conical defect.  From the bulk perspective, the difference between a canonical and microcanonical ensemble is then expressed as a boundary condition: in the former case we fix the size of the Euclidean time circle on the asymptotic boundary, whereas in the latter we naturally fix the Arnowitt-Deser-Misner (ADM) energy~\cite{Arnowitt:1962hi}.  We will see that this leads to nontrivial behaviors of the R\'enyi entropies in a microcanonical ensemble, different from the case of a thermal state.

Finding conical defect solutions with given boundary conditions is a well-defined task, albeit a technically difficult one in general cases without a symmetry.  Here, we will simplify this task by focusing on a particular thermodynamic limit with the entire system size $V\to\infty$ while holding fixed a finite fractional size $f\eq V_A/V$ for $A$ and a finite energy density per ``central charge'' $\e \eq E/CV$ (or, equivalently, a finite temperature $T$ in a thermal state).  Throughout this Letter, we will refer to this limit simply as the \textit{thermodynamic limit with high energy density}.  In this limit, R\'enyi entropies are dominated by a volume-law scaling instead of edge effects near the entangling surface $\pa A$, and we will see that from the bulk perspective this volume-law scaling comes from an approximate planar symmetry which allows us to find the relevant conical defect solutions.  It is worth noting that the volume-law scaling does not necessarily mean that the R\'enyi entropy $S_n^A$ is proportional to the volume $V_A$ (in fact we will see that it is not).  Rather, it means that $S_n^A/V_A$ approaches a well-defined function of the fractional size $f$ in the $V\to\infty$ limit.  For concreteness, we will first take the holographic $C\to\infty$ limit and then the $V\to\infty$ limit.

Our main results in the thermodynamic limit with high energy density are as follows:
\begin{enumerate}
\item In a thermal state at inverse temperature $\b$, the (refined) R\'enyi entropy of $A$ is simply $f$ times the (refined) R\'enyi entropy of the entire system:
\ba
\wtd{S}_n^A(\b) &= f \wtd{S}_n(\b) = f S_\th(n\b), \\
S_n^A(\b) &= f S_n(\b) = \fr{fn\b}{n-1} \[F_\th(n\b) - F_\th(\b)\],
\ea
where $S_\th$ is the thermal entropy and $F_\th$ is the thermal free energy.

\item In a microcanonical ensemble at fixed energy $E$, the (refined) R\'enyi entropy of $A$ is
\ba\la{tsne}
\wtd{S}_n^A(E) &= f S_\th(n\b_n),\\\la{sne}
S_n^A(E) &= \fr{f S_\th(n\b_n) + (1-f) n S_\th(\b_n) - n S_\th(\b_1)}{1-n},
\ea
where $\b_n$ is determined as a function of $n$, $E$, and $f$ by
\be
E = f E_\th(n\b_n) + (1-f) E_\th(\b_n),
\ee
with $E_\th$ the expected total energy in a thermal state.  The R\'enyi entropy~\er{sne} agrees exactly with the proposal of Ref.~\cite{Lu:2017tbo} for individual energy eigenstates.  Here from the holographic perspective, it follows from the much simpler refined R\'enyi entropy~\er{tsne} by using Eq.~\er{sni}.

\end{enumerate}

\se{Thermal State}

In a QFT, a thermal state $e^{-\b H}/\tr(e^{-\b H})$ on a spatial manifold $\S$ is described by a Euclidean path integral on $\S \times S^1_\b$, where $S^1_\b$ denotes the Euclidean time circle of size $\b$.  In gauge-gravity duality, the corresponding bulk geometry is a smooth solution to the bulk equations of motion with $\S \times S^1_\b$ as the conformal boundary.  Let us call this bulk geometry $\cB(\b)$.

\begin{figure}[t]
\centering
\includegraphics[width=\columnwidth]{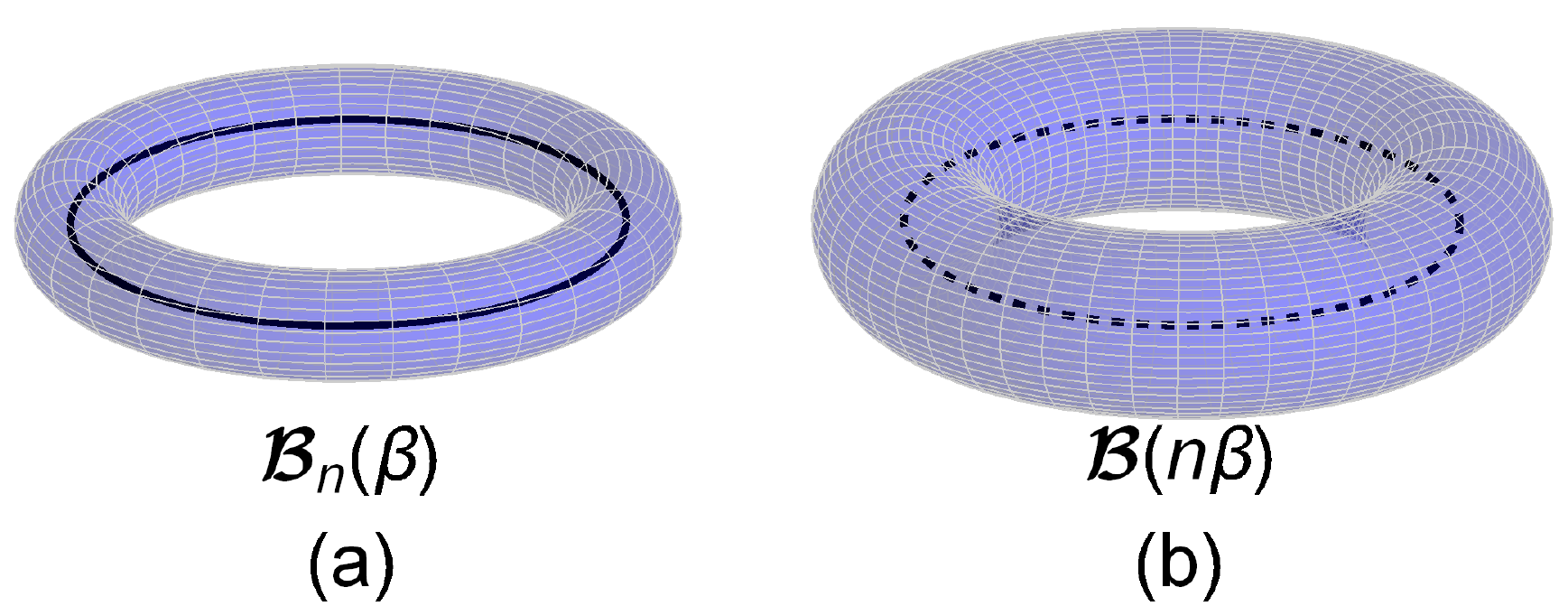}
\caption{(a) Bulk geometry $\cB_n(\b)$ with a cosmic brane (black solid circle) homologous to the boundary space $\S$ (a circle here) in a thermal state at inverse temperature $\b$. (b) Bulk geometry $\cB(n\b)$ with a smooth Euclidean black hole horizon (black dotted circle) in a thermal state at inverse temperature $n\b$.  The left geometry $\cB_n(\b)$ can be obtained as a $\bZ_n$ quotient of the right geometry $\cB(n\b)$, with the cosmic brane identified with the black hole horizon.}
\la{entire}
\end{figure}

In the high temperature limit that we are interested in, the bulk geometry $\cB(\b)$ describes a large Euclidean black hole.  This is a solution in which the Euclidean time circle $S^1_\b$ contracts to zero size smoothly at the horizon (Fig.~\re{entire}).  The horizon area determines the thermal entropy $S_\th(\b)$ of the entire system~\cite{Bekenstein:1973ur,Bardeen:1973gs,Hawking:1974sw,Gibbons:1976ue,Ryu:2006bv,Ryu:2006ef}.

Let us first study the R\'enyi entropies of the entire system.  The holographic prescription~\er{tsnb} asks for a bulk solution with a conical defect homologous to the entire $\S$ with a conical opening angle $2\pi/n$.  Let us call this conical geometry $\cB_n(\b)$ and the conical defect in it $\cC_{n,\b}$.  In fact, $\cB_n(\b)$ is easily obtained as a quotient of the smooth geometry $\cB(n\b)$:
\be
\cB_n(\b) = \cB(n\b) / \bZ_n,
\ee
where the $\bZ_n$ group is generated by a shift in the Euclidean time coordinate by $\b$.  The black hole horizon in $\cB(n\b)$ is precisely the set of fixed points under the $\bZ_n$ action and becomes the conical defect in $\cB_n(\b)$ after the $\bZ_n$ quotient (Fig.~\re{entire}).  Therefore, the area of the cosmic brane $\cC_{n,\b}$ in $\cB_n(\b)$ is equal to the horizon area in $\cB(n\b)$, and Eq.~\er{tsnb} implies
\be\la{tst}
\wtd{S}_n(\b) = \fr{\Area(\cC_{n,\b})}{4G_N} = S_\th(n\b).
\ee
This agrees with the general property that the refined R\'enyi entropy $\wtd{S}_n$ can be defined alternatively as the von Neumann entropy of the density matrix $\r^n / \tr \r^n$.

\begin{figure}[t]
\centering
\includegraphics[width=\columnwidth]{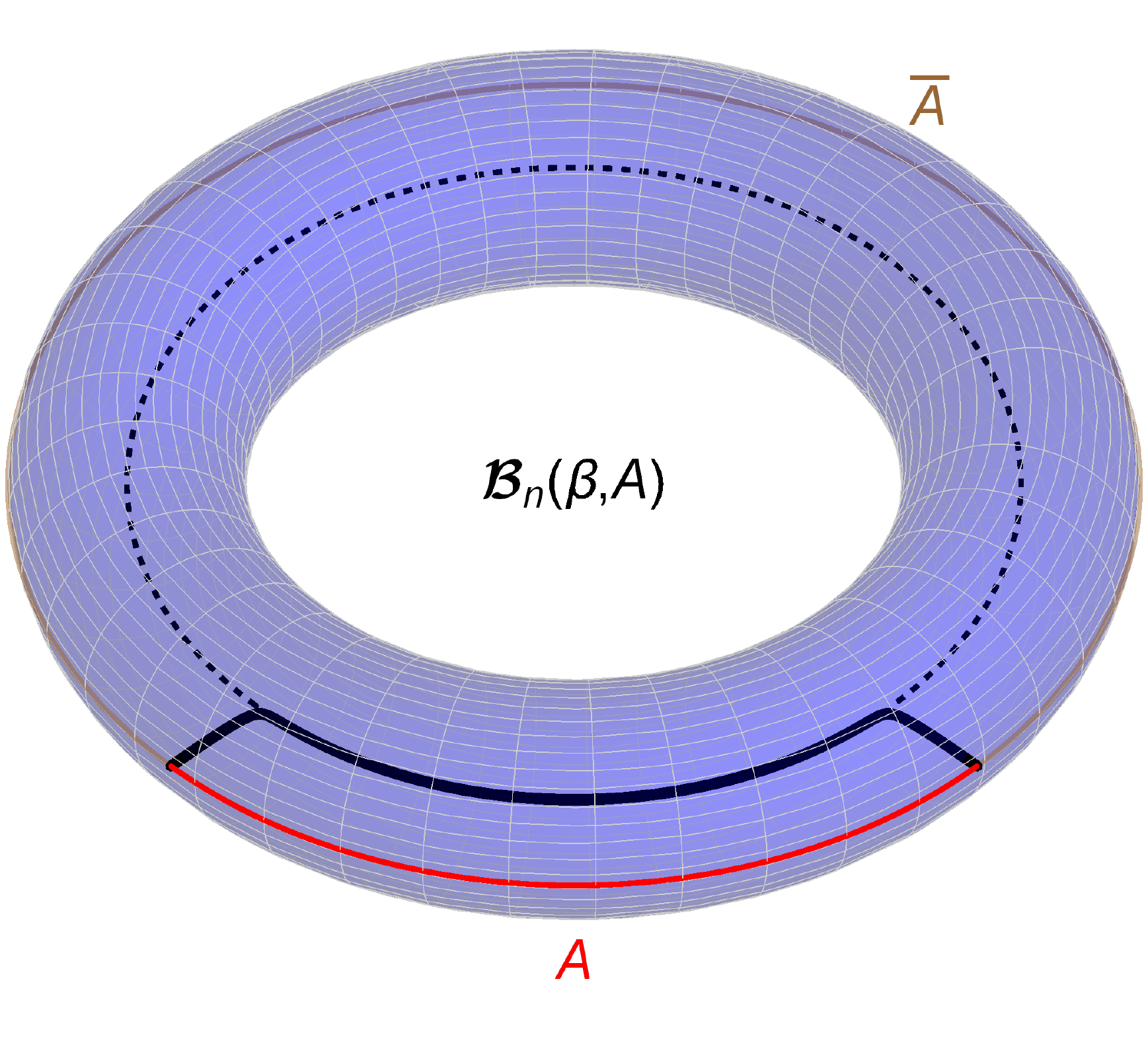}
\caption{Bulk geometry $\cB_n(\b,A)$ with a cosmic brane (black solid curve) homologous to a subregion $A$ (red interval) in a thermal state at inverse temperature $\b$.  Up to edge effects near $\pa A$, this geometry can be approximately constructed by gluing the segment of the $\cB_n(\b)$ geometry corresponding to $A$ and the segment of $\cB(\b)$ corresponding to $\ol{A}$.}
\la{region}
\end{figure}

Now consider the R\'enyi entropy of a subregion $A$.  According to Eq.~\er{tsnb}, we need a bulk solution with a conical defect homologous to $A$.  Let us call this geometry $\cB_n(\b,A)$.  It is not easy to construct this geometry exactly, but in the high temperature, thermodynamic limit where the R\'enyi entropy is dominated by a volume-law scaling instead of edge effects near $\pa A$, the geometry $\cB_n(\b,A)$ is approximately given by $\cB_n(\b)$ within the bulk subregion $x \in A$ and by $\cB(\b)$ within the subregion $x \in \ol{A}$, where $x$ denotes a collection of coordinates describing directions parallel to $\S$ (Fig.~\re{region}).  In other words, the geometries $\cB_n(\b)$ and $\cB(\b)$ enjoy an approximate planar symmetry in the $x$ directions when we take the high temperature, thermodynamic limit, and the wanted geometry $\cB_n(\b,A)$ can be approximated by gluing the part of $\cB_n(\b)$ corresponding to $A$ and the part of $\cB(\b)$ corresponding to $\ol{A}$.

According to Eq.~\er{tsnb}, the refined R\'enyi entropy $\wtd{S}_n^A$ in a thermal state is given by the cosmic brane area in $\cB_n(\b,A)$, or $f$ times the cosmic brane area~\er{tst} in $\cB_n(\b)$:
\be\la{tsnt}
\wtd{S}_n^A(\b) = f \wtd{S}_n(\b) = f S_\th(n\b).
\ee
Plugging this into Eq.~\er{sni}, we find the R\'enyi entropy
\be\la{snt}
S_n^A(\b) = f S_n(\b) = \fr{fn\b}{n-1} \[F_\th(n\b) - F_\th(\b)\].
\ee

Even though our results do not require the boundary QFT to have additional symmetries, it is worth considering the special case where it is a $d$-dimensional conformal field theory (CFT), in which the thermal entropy is determined by scale invariance up to a coefficient:
\be\la{scft}
S_\th(\b) = \fr{C V}{\b^{d-1}}.
\ee
Using this, we find that Eqs.~\er{tsnt} and~\er{snt} become
\ba
\wtd{S}_n^A(\b) &= \fr{f C V}{(n\b)^{d-1}}, \\
S_n^A(\b) &= \fr{f C V}{(n\b)^{d-1}} \fr{n^d-1}{d(n-1)}.
\ea

\se{Microcanonical ensemble}

In a microcanonical ensemble, the state of the entire system is characterized by a fixed total energy $E$ (with some energy width) instead of a fixed Euclidean time circle size $\b$.  Therefore, when applying the holographic prescription~\er{tsnb} for R\'enyi entropies we need to find a bulk conical solution with the asymptotic boundary condition fixing the ADM energy to be $E$.  This prescription can be derived by applying the replica method of Ref.~\cite{Dong:2016fnf} in a microcanonical path integral that fixes the total energy; such microcanonical path integrals were recently studied in Ref.~\cite{Marolf:2018ldl} (which was partly motivated by Ref.~\cite{Brown:1992bq}, although there the energy density at each point is fixed instead of just the total energy).

We now show that the required bulk solution is precisely $\cB_n(\b_n,A)$ defined earlier with some particular value of $\b_n$ that depends on $n$ as well as implicitly on $E$ and $f$.  To see this, we note that the geometry $\cB_n(\b_n,A)$ is indeed a bulk solution that has a conical defect homologous to $A$ with the correct conical opening angle $2\pi/n$.  The only remaining condition for $\cB_n(\b_n,A)$ to be the correct solution is that its ADM energy should be $E$.  In the thermodynamic limit with high energy density, $\cB_n(\b_n,A)$ is approximately given by $\cB_n(\b_n)$ for $x \in A$ and by $\cB(\b_n)$ for $x \in \ol{A}$ as mentioned earlier.  The ADM energy can be calculated by integrating over $\S$ the strength of the gravitational field at the asymptotic boundary.  Therefore, the ADM energy of $\cB_n(\b_n,A)$ is a weighted average of the ADM energies of $\cB_n(\b_n)$ and $\cB(\b_n)$ which we fix to be $E$:
\be\la{efix}
E = f E_\th(n\b_n) + (1-f) E_\th(\b_n).
\ee
Here we have used that the ADM energy of $\cB_n(\b_n)$ is the same as that of $\cB(n\b_n)$ and equal to the expected energy $E_\th(n\b_n)$ in the corresponding thermal state.  Eq.~\er{efix} uniquely determines $\b_n$ as a function of $n$, $E$, and $f$.

According to Eq.~\er{tsnb}, the refined R\'enyi entropy $\wtd{S}_n^A$ in a microcanonical ensemble is determined by the cosmic brane area in $\cB_n(\b_n,A)$, which is precisely given by Eq.~\er{tsnt} with $\b$ set to $\b_n$:
\be\la{tsne2}
\wtd{S}_n^A(E) = f S_\th(n\b_n).
\ee
Again, we determine the R\'enyi entropy from this by using Eq.~\er{sni}.  The integral with respect to $n$ is nontrivial due to the dependence of $\b_n$ on $n$, but a closed-form expression for the R\'enyi entropy can be obtained,
\be\la{sne2}
S_n^A(E) = \fr{f S_\th(n\b_n) + (1-f) n S_\th(\b_n) - n S_\th(\b_1)}{1-n},
\ee
in terms of $\b_n$ which is defined by Eq.~\er{efix}.

The microcanonical R\'enyi entropy~\er{sne2} is different from the much simpler R\'enyi entropy~\er{snt} in a thermal state, showing that the entanglement spectrum of a subsystem of finite fractional size in a microcanonical ensemble significantly deviates from a purely thermal behavior.  In particular, the microcanonical R\'enyi entropy~\er{sne2} is not linear in $f$ due to the dependence of $\b_n$ on $f$: for $n>1$ ($n<1$), $\b_n$ decreases (increases) with $f$ according to Eq.~\er{efix}, and Eq.~\er{sne2} gives a concave (convex) function of $f$ with $\pa^2 S_n^A(E)/\pa f^2>0$ ($<0$)~\cite{Lu:2017tbo}.  On the other hand, the microcanonical R\'enyi entropy~\er{sne2} approaches the thermal case~\er{snt} at inverse temperature $\b_1$ in two special limits.  One is the $f \to 0$ limit keeping terms that are linear in $f$, in agreement with the derivation of the canonical ensemble for small subsystems from the microcanonical ensemble.  The other limit is $n \to 1$, showing that the von Neumann entropy of a subsystem cannot distinguish microcanonical ensembles from canonical ensembles.  This is in contrast to the R\'enyi entropies which are more sensitive to the fine-grained details of a quantum state.

In the special case of a $d$-dimensional CFT, we can use Eq.~\er{scft} to find $E_\th(\b) = (d-1) C V / (d \b^d)$ and solve Eq.~\er{efix}:
\be
\b_n = \[\fr{(d-1) C V \(1-f+f n^{-d}\)}{d E}\]^{1/d}.
\ee
Plugging this into Eqs.~\er{tsne2} and~\er{sne2}, we find
\ba
\wtd{S}_n^A(E) &= f S_\mc(E) n^{1-d} \(1-f+f n^{-d}\)^{(1-d)/d}, \\\la{sncft}
S_n^A(E) &= \fr{n S_\mc(E)}{n-1} \[1- \(1-f+f n^{-d}\)^{1/d}\],
\ea
where the microcanonical entropy $S_\mc(E)$ is given by
\be
S_\mc(E) = C V \(\fr{d \e}{d-1}\)^{(d-1)/d},
\ee
with $\e \eq E/(CV)$.  The R\'enyi entropy~\er{sncft} again agrees exactly with the proposal of Ref.~\cite{Lu:2017tbo} for individual energy eigenstates.

\se{Discussion}

We have shown holographically that R\'enyi entropies of any subsystem of finite fractional size can be used to diagnose the state of the entire system.  In particular, they differ between a microcanonical and a canonical ensemble for the entire system.  This is in sharp contrast to the von Neumann entropy which is more coarse-grained and cannot distinguish these two types of ensembles.  From the holographic perspective, this is because R\'enyi entropies are determined from bulk cosmic branes which lead to different backreacted geometries due to distinct boundary conditions in these two types of ensembles, whereas the von Neumann entropy is determined from extremal surfaces that only see the same semiclassical geometry.  This distinction between R\'enyi entropies and the von Neumann entropy is reminiscent of a thermodynamic explanation of a similar phenomenon given in Ref.~\cite{Headrick:2013zda}.

The holographic prescription~\er{tsnb} used to calculate the R\'enyi entropies assumes that the boundary QFT is dual to a bulk theory described by Einstein gravity, but none of our results are modified in more general cases where the bulk theory includes higher-derivative corrections.  In these cases, we replace the area of the cosmic brane in Eq.~\er{tsnb} by some generalized notion of area~\cite{Dong:2013qoa,Camps:2013zua,Dong:2017xht}.  The form of this generalized area is completely determined by the bulk action, but it is sufficient for our purposes to know that the generalized area is an integral over the conical defect of some combination of local geometric invariants.  Since our calculation only relies on this locality, the results remain unchanged.

Our results have been obtained in the large $C$ limit of the boundary QFT where the bulk physics is dominated by classical solutions, but again these results are not modified by quantum corrections (at least perturbatively).  This is because these quantum corrections take the form of the refined R\'enyi entropies in the bulk entanglement wedge~\cite{Faulkner:2013ana,Engelhardt:2014gca,Dong:2017xht}, and they scale with the cosmic brane area (and therefore do not affect our calculation) in the thermodynamic limit with high energy density.

For simplicity we have worked in the thermodynamic limit with high energy density so that R\'enyi entropies satisfy a volume-law scaling in the sense that $S_n^A/V_A$ approaches a well-defined limit.  However, our results actually hold in a larger regime defined by only requiring the volume-law term to dominate over edge effects (mainly an area-law term) near the entangling surface $\pa A$.  In this larger regime, $S_n^A$ is well approximated by our results even though it does not necessarily scale with the volume.  In the special case of a $d$-dimensional CFT, the larger regime is $E \gg C V^{1-d/(d-1)^2} \d^{-d(d-2)/(d-1)}$ for a microcanonical ensemble or $T \gg V^{-1/(d-1)^2} \d^{-(d-2)/(d-1)}$ for a canonical ensemble, with $\d$ the short-distance cutoff (in $d=2$, the powers of $\d$ are replaced by $\ln\d$).

Even though we have focused in the previous section on a microcanonical ensemble at fixed energy $E$ (with some energy width), our results~\er{tsne2} and~\er{sne2} could also apply to an energy eigenstate $|E\>$ when the subsystem is no larger than half of the entire system (as proposed in Ref.~\cite{Lu:2017tbo}).  Here we provide an intuitive, holographic argument for this.  In the case of an energy eigenstate, the cosmic brane prescription~\er{tsnb} for R\'enyi entropies requires a small modification: instead of requiring the cosmic brane $\cC$ to be homologous to the subregion $A$, we only need it to be anchored on the boundary of $A$.  Whereas previously the homology constraint requires the existence of a bulk codimension-1 surface $\cE$ whose boundary is $A \cup \cC$, now we only need the boundary of $\cE$ to be $A \cup \cC$ up to horizons of pure-state black holes.  This modified prescription with a weakened homology constraint is consistent in the sense that it gives the correct, vanishing R\'enyi entropies for the entire system, as well as the correct von Neumann entropy in two-dimensional holographic CFTs~\cite{Asplund:2014coa,Almheiri:2016blp}.  It is also reminiscent of the interpretation of the homology constraint given in Refs.~\cite{Almheiri:2016blp,Harlow:2016vwg}.  Whether this modified prescription can be firmly established for holographic R\'enyi entropies in general energy eigenstates is an open question that we leave to the future.  We would like to mention, however, that two-dimensional CFTs are likely special in this regard, and a different expression for R\'enyi entropies in Virasoro primary states in the large central charge limit was already obtained in Ref.~\cite{Faulkner:2017hll}.  This could be explained by the fact that two-dimensional CFTs have an infinite number of commuting conserved charges known as the quantum Korteweg-de Vries (KdV) charges~\cite{Bazhanov:1994ft}, leading to interesting thermodynamic properties~\cite{deBoer:2016bov,Basu:2017kzo,He:2017txy,Lashkari:2017hwq,Dymarsky:2018lhf,Maloney:2018hdg,Maloney:2018yrz,Dymarsky:2018iwx} and differences between primary and descendant states.  We leave a full resolution of this puzzle to future work.

\se{Acknowledgments}

I thank Anatoly Dymarsky, Thomas Faulkner, Tarun Grover, Donald Marolf, Henry Maxfield, and Huajia Wang for useful discussions.  This work was supported in part by the U.S. Department of Energy under Grant No.\ DE-SC0019139 and by funds from the University of California.  I am also grateful to the KITP for hospitality during part of the development of this work.  The KITP was supported in part by the National Science Foundation under Grant No.\ PHY-1748958.

\bibliography{bibliography}

\end{document}